\documentclass[12pt,preprint]{aastex}
\usepackage{textcomp}
\usepackage{amsmath}
\usepackage{graphicx}
\usepackage[T1]{fontenc}

\def\beq{\begin{eqnarray}}
\def\eeq{\end{eqnarray}}

\begin{document}

\title{Evolutions of stellar-mass black hole hyperaccretion systems in the center of gamma-ray bursts}

\author{Cui-Ying Song\altaffilmark{1,2}, Tong Liu\altaffilmark{1,3}, Wei-Min Gu\altaffilmark{1}, Shu-Jin Hou\altaffilmark{4}, Jian-Xiang Tian\altaffilmark{2}, and Ju-Fu Lu\altaffilmark{1}}

\altaffiltext{1}{Department of Astronomy, Xiamen University, Xiamen, Fujian 361005, China; tongliu@xmu.edu.cn}
\altaffiltext{2}{Shandong Provincial Key Laboratory of Laser Polarization and Information Technology, Department of Physics, Qufu Normal University, Jining, Shandong 273165, China; jxtian@dlut.edu.cn}
\altaffiltext{3}{Department of Physics and Astronomy, University of Nevada, Las Vegas, NV 89154, USA}
\altaffiltext{4}{College of Physics and Electronic Engineering, Nanyang Normal University, Nanyang, Henan 473061, China}

\begin{abstract}
A neutrino-dominated accretion disk around a stellar-mass black hole (BH) can power a gamma-ray burst (GRB) via annihilation of neutrinos launched from the disk. For the BH hyperaccretion system, high accretion rate should trigger the violent evolution of the BH's characteristics, which further leads to the evolution of the neutrino annihilation luminosity. In this paper, we consider the evolution of the accretion system to analyze the mean time-dependent neutrino annihilation luminosity with the different mean accretion rates and initial BH parameters. By time-integrating the luminosity, the total neutrino annihilation energy with the reasonable initial disk mass can satisfy the most of short-duration GRBs and about half of long-duration GRBs. Moreover, the extreme Kerr BH should exist in the cental engines of some high-luminosity GRBs. GRBs with higher energy have to request the alternative magnetohydrodynamics processes in the centers, such as the Blandford-Znajek jet from the accretion system or the millisecond magnetar.
\end{abstract}


\section{Introduction}

Gamma-ray bursts (GRBs), known for more than 40 yr \citep{Klebesadel1973}, are dramatic flashes of gamma rays in the universe. The timescales of GRBs are from a fraction of a second up to a few hundred seconds, which are grouped into two classes by their characteristic duration $T_{90}$ \citep{Kouveliotou1993}. Long-duration GRBs ($T_{90} > 2~\rm s$, hereafter LGRBs) are widely thought to come from a gravitational collapse of a massive star \citep[e.g.,][]{Woosley1993}, and short-duration GRBs ($T_{90} < 2~\rm s$, hereafter SGRBs) are related to merger events of a neutron star (NS) binary or an NS-black hole (BH) binary \citep[e.g.,][]{Eichler1989,Narayan1992,Nakar2007}. All of the above scenarios may lead to the formation of a hyperaccretion disk around a stellar-mass BH \citep[e.g.,][]{Popham1999,Gu2006,Liu2007} or a magnetar \citep[e.g.,][]{Usov1992,Dai2006,Gao2013b,Yu2013}.

In the accretion scenarios, the Blandford-Znajek (BZ) mechanism \citep{Blandford1977} and neutrino annihilation process are generally considered to power GRBs \citep[e.g.,][]{Liu2015b}. For the BZ mechanism, the sufficiently strong magnetic fields are bounded near the BH horizon; thus, a Poyting jet can be launched to trigger GRBs. Moreover, a neutrino-dominated accretion flow (NDAF) around a stellar-mass BH with an extremely high accretion rate is also a popular candidate of the central engine of GRBs \citep[e.g.,][]{Popham1999,Narayan2001,DiMatteo2002,Kohri2002,Rosswog2002,Kohri2005,Lee2005,Gu2006,Chen2007,Kawanaka2007,Liu2007,Liu2010,Liu2012a,Liu2013, Liu2015a,Lei2009,Zalamea2011,Janiuk2013,Kawanaka2013,Li2013,Xue2013}. For the NDAF model, the matter of the disk is too dense and hot to be cooled via the photon emission. Instead, neutrinos tap the thermal energy of the disk produced by the viscous dissipation and liberate tremendous amounts of binding energy, and their annihilation above the disk can produce the original fireball.

In the NDAF model, the neutrino radiation and annihilation luminosity are closely related to the accretion rate and the mass and spin of the BH \citep[see, e.g.,][]{Fryer1999,Popham1999,Liu2007,Zalamea2011,Xue2013}. The BH evolution may be intense in this hyperaccretion system, and it will influence on the neutrino annihilation luminosity.
\citet{Leng2014} derived the maximum possible energy of a neutrino-powered jet as a function of the burst duration. Their results demonstrated that the model can potentially power GRBs with the durations less than 100 s, so the progenitors of the late X-ray flares and ultralong GRBs are difficult to explain by the model. Unfortunately, they ignored the energy of afterglow coming from the activities of the central engine and the evolution of the BH, especially that of the BH spin. \citet{Liu2015a} collected the current SGRBs observational data and estimated the disk masses for the different characteristics of the BH in the framework of the fireball and neutrino annihilation model. The results showed that the disk mass of a certain SGRB mainly depends on its output energy, jet opening angle, and BH characteristics. Even for the extreme BH parameters, some high-energy SGRBs still require the massive disks. Nonetheless, the evolution of the central BH has been neglected. If an NDAF around a stellar-mass BH really exists in the center of a GRB, a violent evolution of the BH's characteristics should inevitably occur, which leads to the evolution of the neutrino annihilation luminosity. Thus, it is necessary to take the time evolution of the system into account in the calculations of the luminosity and examine its feasibilities compared with the observations.

This paper is organized as follows. In Section 2, we describe the analytical model for the evolutive central engine of GRBs. Numerical results and comparisons with the observations are shown in Section 3. Conclusions and discussion are presented in Section 4.

\section{Model}

If there is a hyperaccretion system in the center of a GRB, the characteristic parameters of the BH will be significantly time dependent. Since the terms relevant for the BZ mechanism are not included, the evolution equations of a Kerr BH, based on the conservation of energy and angular momentum, can be expressed by \citep[e.g.,][]{Liu2012b}
\beq
\frac{d M_{\rm BH}}{dt}=\dot{M}e_{\rm ms},
\eeq
\beq
\frac{d J_{\rm BH}}{dt}=\dot{M}l_{\rm ms},
\eeq
where $M_{\rm BH}$, $J_{\rm BH}$, and $\dot{M}$ are the mass and angular momentum of the BH and the mean mass accretion rate, respectively. Parameters $m_{\rm BH}=M_{\rm BH}/M_{\odot}$ and $\dot{m}=\dot{M}/(M_{\odot}~\rm s^{-1})$ are widely introduced in the following solutions. The specific energy and angular momentum corresponding to the marginally stable orbit radius $r_{\rm ms}$ of the disk, i.e., $e_{\rm ms}$ and $l_{\rm ms}$, can be written as \citep[e.g.,][]{Novikov1973,Wu2013,Hou2014}
\beq
e_{\rm ms}= \frac{1}{\sqrt{3 x_{\rm ms}}} (4- \frac{3 a_\ast}{\sqrt{x_{\rm ms}}}),
\eeq
\beq
l_{\rm ms}=2 \sqrt{3} \frac{G M_{\rm BH}}{c} (1-\frac{2 a_\ast}{3\sqrt{x_{\rm ms}}}),
\eeq
where $a_*\equiv cJ_{\rm BH}/GM_{\rm BH}^{2}$ is the dimensionless spin parameter of the BH and $x_{\rm ms}=3+Z_{2}-\sqrt{(3-Z_{1})(3+Z_{1}+2Z_{2})}$ is the dimensionless marginally stable orbit radius of the disk \citep[e.g.,][]{Bardeen1972,Kato2008}, where $Z_{1}=1+(1-a_*^{2})^{1/3}[(1+a_*)^{1/3}+(1-a_*)^{1/3}]$ and $Z_{2}=\sqrt{3a_*^{2}+Z_{1}^{2}}$ for $0 < a_* <1$.

Therefore, by incorporating Equations (1)-(4), the evolution of the BH spin is expressed by
\beq
\frac{d{a}_\ast}{dt} =  2\sqrt{3} \frac{\dot{M}}{M_{\rm BH}} (1-\frac{a_\ast}{\sqrt{x_{\rm ms}}})^2.
\eeq
If the initial mass and spin of the BH, $M_i$ and $a_i$, are given, according to the above equations, we can obtain the characteristics of the BH at any time.

The disk mass can be estimated by
\beq
M_{\rm disk}=\dot{M}~\frac{T_{\rm 90}}{1+z},
\eeq
where $T_{\rm 90}$ and $z$ are the burst duration and red shift, respectively. For simplicity, we also use $m_i=M_i/M_\odot$ and $m_{\rm disk}=M_{\rm disk}/M_\odot$ below.

In the NDAF model, since the neutrino annihilation luminosity $L_{\nu\bar{\nu}}$ is related to $m_{\rm BH} $, $a_*$, and $\dot{m}$, a violent evolution of the BH's mass and angular momentum will lead to the evolution of neutrino annihilation luminosity. Here we use the approximate analytical formula of the luminosity $L_{\nu\bar{\nu}}$ in \citet{Zalamea2011}, which is written as
\beq
L_{\nu\bar{\nu}}&\approx& 1.59 \times 10^{54} ~x_{\rm ms}^{-4.8}~m_{_{\rm BH}}^{-3/2} \nonumber \\
&&\times \Bigg \{
\begin{array}{ll}
0 & \hbox{for $\dot{m} < \dot {m}_{\rm ign}$}\\
\dot{m }^{9/4} & \hbox{for $\dot {m}_{\rm ign} < \dot{m} < \dot{m}_{\rm trap}$}\\
\dot{m}_{\rm trap}^{9/4} & \hbox{for $\dot{m} > \dot{m}_{\rm trap}$}\\
\end{array}
\Bigg \} ~\rm erg~s^{-1}.
\eeq
The dimensionless characteristic accretion rates $\dot {m}_{\rm ign}$ and $\dot {m}_{\rm trap}$ \citep[e.g.,][]{DiMatteo2002,Kohri2005,Chen2007,Liu2012a,Xue2013} depend on the viscosity parameter $\alpha$ and $a_*$ in the numerical calculations. We checked that the accretion rates in our results are in the suitable ranges $\dot {m}_{\rm ign} < \dot{m} < \dot{m}_{\rm trap}$. Because the value of the viscosity parameter has little effect on $L_{\nu\bar{\nu}}$, $\alpha = 0.1$ is adopted here \citep[e.g.,][]{Zalamea2011,Liu2015a}. Furthermore, the total neutrino annihilation energy can be obtained by integrating Equation (7),
\beq
E_{\nu\bar{\nu}}=1.59 \times 10^{54} ~\int_0^{T_{90,\rm s}/(1+z)} x_{\rm ms}^{-4.8}~m_{_{\rm BH}}^{-3/2}~\dot{m}^{9/4}d t_{\rm s} ~\rm ergs,
\eeq
where $T_{90, \rm s}=T_{90}/(1~\rm s)$ and $t_{\rm s}=t/(1~\rm s)$.

The energy of the previous fireball to power the prompt emission and afterglow of GRBs is deposited predominantly via neutrino and antineutrino annihilation in the polar funnel region \citep[e.g.,][]{Narayan1992,DiMatteo2002,Liu2007}. Thus, the neutrino annihilation energy corresponds to the sum of the isotropic radiated energy from the observational data in the prompt emission phase E$_{\rm \gamma,iso}$ and the isotropic kinetic energy of the outflow powering long-lasting afterglow $E_{\rm k,iso}$ \citep[e.g.,][]{Fan2011,Liu2015c}:
\beq
E_{\nu\bar{\nu}}=(E_{\rm \gamma,iso}+E_{\rm k,iso}) \theta_{\rm j}^{2}/\eta,
\eeq
where $\eta$ is the conversion factor; we adopt $\eta=0.3$ here \citep[e.g.,][]{Aloy2005}.

To convert the measured jet break times $t_{\rm j}$ in the X-ray afterglow phase of GRBs to opening angles $\theta_{\rm j}$ of the conical blast wave, we used the formulation of \citet{Frail2001}:
\beq
\theta_{\rm j} \approx 0.076~(\frac{t_{\rm j}}{1~{\rm day}})^{3/8}(\frac{{1+z}}{2})^{-3/8}(\frac{n}{\rm 0.01~cm^{-3}})^{1/8} (\frac{E_{\rm k,iso}}{10^{51}~\rm ergs})^{-1/8},
\eeq
where $n$ is the number density of the burst circumstance.

E$_{\rm \gamma,iso}$ can be calculated by the observational data, which is defined as \citep[e.g.,][]{Liu2015b,Liu2015c}
\beq
E_{\rm \gamma,iso}=4\pi D_L^2 F_\gamma /(1+z),
\eeq
where $D_{\rm L}$ is the luminosity distance and $F_\gamma$ is the fluence in the 15-150 keV for $\emph{Swift}$ events.
$E_{\rm k,iso}$ can be deduced from the modeling of the X-ray afterglow data, which is written as \citep{Lloyd-Ronning2004,Fan2006a,Zhang2007,Liu2015c}
\beq
E_{\rm k,iso} &\approx& 9.2 \times 10^{52} R L_{X,46}^{4/(p+2)}(\frac{1+z}{2})^{-1} \epsilon_{B,-2}^{(2-p)/(p+2)} \epsilon_{\rm e,-1}^{4(1-p)/(p+2)}\nonumber\\ &&\times t_{\rm d}^{(3p-2)/(p+2)} (1+Y)^{4/(p+2)}~{\rm ergs},
\eeq
where $R \sim (t_{11}/T_{90, \rm s})^{17\epsilon_e/16}$ is a factor that accounts for the energy loss during the deceleration following the prompt gamma-ray emission phase \citep[e.g.,][]{Sari1997,Lloyd-Ronning2004}, $L_{X,46}={L_X/(10^{46}}~\rm erg~s^{-1})$ is the isotropic X-ray afterglow luminosity, $\epsilon_{\rm e,-1}=\epsilon_{\rm e}/0.1$ is the fraction of shock energy given to the electrons, $\epsilon_{\rm B,-2}=\epsilon_{\rm B}/0.01$ is the fraction of energy in the magnetic field, $t_{11}=t/({\rm 11~hours})$ and $t_{\rm d}=t/(\rm 1~day)$ are the time of observation, $Y$ is the Compton parameter, and $p$ is the energy distribution index of the shock-accelerated electrons and can be fitted by the observed photon index in the X-ray spectrum \citep[e.g.,][]{Zhang2006,Gao2013a}.

\section{Results}

According to the above equations, we can describe the evolution of the neutrino annihilation luminosity by setting the initial mass and spin of the BH. Furthermore, we can also calculate the total neutrino annihilation energy to measure the ability of this mechanism by comparing with the GRB data.

Additionally, it is found that the value of the disk mass is the key factor to reflect the rationality of our model. SGRBs originate from the merger events of the compact objects. In this case, the BH mass is naturally less than the total mass of the binary, i.e., $< 4~M_\odot$, and the disk mass is about $0.2$ or $0.5~M_\odot$ for the mergers of two NSs or a BH-NS binary, which is proposed in further calculations and simulations \citep[e.g,][]{Kluzniak1998,Ruffert1998,Lee1999,Popham1999,Liu2012b,Fryer2014}. In the LGRB case, the newborn BH originates from its progenitor star, and its mass should also be a stellar-mass order \citep[e.g.,][]{Heger2000} and the disk mass was about several solar mass \citep[e.g.,][]{Popham1999}. It is reasonable to assume that $m_i=2.3,~3$ and $3,~4$, and $m_{\rm disk}=0.2,~0.5$ and $1,~3,~5$ for the cases of SGRBs and LGRBs in our calculations, respectively. Moreover, the BH spin parameters are adopted as $a_i=0.5, 0.9,$ and $0.99$. The results for SGRBs and LGRBs are discussed as follows.

\subsection{SGRBs}

Figure~1 shows the evolution of the mean neutrino annihilation luminosity for SGRBs. The black, red, and blue lines show the profiles with an initial spin of BH $a_i=0.5, 0.9,$ and $0.99$, respectively. The solid and dashed lines correspond to the dimensionless mean accretion rate $\dot{m}=0.2$ and $0.5$ with $m_{i}=2.3$, apart from the dot-dashed line with initial dimensionless BH mass $m_{i}=3$. The dimensionless disk masses $m_{\rm disk}$ are equal to $0.2$ and $0.5$, which are marked with triangles and squares, respectively. The maximum of the mean neutrino annihilation luminosity for the blue dashed line is $3.05\times10^{52} \rm ~erg~s^{-1}$, and the break indicates that the BH spin has evolved to $0.998$ and its evolution has stopped \citep[e.g.,][]{Kato2008}. According to the comparison of the six lines, we can find that the influence of the BH spins and mean accretion rates is more significant on $L_{\nu\bar{\nu}}$ than these of the BH masses. Thus, the single value of $m_i$ is adopted in the following figures of SGRBs. Here the mean neutrino annihilation luminosity calculated with a mean accretion rate is not directly associated with observational data. Furthermore, we adopt the Keplerian angular velocity in the inner radius of the disk, while a more possible angular velocity is sub-Keplerian, as well as that in the advection-dominated accretion flow (ADAF) model \citep[e.g.,][]{Narayan1997}. If the latter form is true, we can expect that the maximum of the mean neutrino annihilation luminosity will be smaller than the present peak value and the model will be more powerless to explain SGRBs.

In order to definitely examine the evolutive neutrino annihilation model, firstly, we adopt $E_{\rm \gamma,iso}$, $E_{\rm k,iso}$, $\theta_{\rm j}$, and other data of 30 SGRBs shown in Table 1 of \citet{Liu2015c} to calculate their individual neutrino annihilation energy and then to compare with the typical theoretical lines. As shown in Figure 2, all parameters presented by the colors and styles of lines are the same as in Figure 1. All curves correspond to the initial value of BH mass $m_i=2.3$ and typical red shift $z=0.5$. While the disk mass $M_{\rm disk}=0.5 ~M_{\odot}$, all lines are truncated. The magenta points show the total neutrino annihilation energy in the duration $T_{90}$ calculated by observational data. We notice that all points are under our predicted lines, which means that our model can interpret the profiles in SGRB samples. Moreover, the high BH spin or the massive disk is required for some high-luminosity or long-duration SGRBs.

We further tested the consequences of this model by studying the distribution of the disk mass. In Figure 3, we estimate the disk mass with the given BH initial mass and spin for the sample of 30 SGRBs. We note that the disk mass of most SGRBs is less than $0.4~M_{\odot}$, and in some special cases it is larger than the limits. Even for the case of Figure 3(b), there still exists one SGRB whose disk mass is larger than $0.4~M_{\odot}$. It implies that the more extreme spin parameter or another energy mechanism may exist in the center. The results of Figure 3 are generally in agreement with \citet{Liu2015c}.

\subsection{LGRBs}

We also plot the time evolution of the mean neutrino annihilation luminosity for LGRBs as shown in Figure 4. The black solid and dashed lines correspond to the BH spin parameter $a_i=0.5$ and $0.9$ with the dimensionless mean accretion rate $\dot{m}=0.1$ and $m_i=3$. The red solid and dashed lines correspond to $m_{i}=3$ and $4$ with $a_i=0.9$, and $\dot{m}=0.2$. $m_{\rm disk}=1, 3,$ and $5$ are marked with the triangles, squares, and five-pointed stars, respectively. The maximum of the luminosity of four lines from top to bottom is $18.4,~11.9,~3.86,$ and $2.46 \times 10^{50} ~\rm erg~s^{-1}$. For the long-duration accretion processes, the breaks appear in all cases. Furthermore, similar to SGRBs, we focus on the effects of the initial spins and mean accretion rates on the mean neutrino annihilation luminosity, and we set $m_{i}=3$ in the following figures. Similarly, if we take the sub-Keplerian form, the mean neutrino annihilation will be more powerless to explain LGRBs.

Figure 5 displays the predictions of our model compared with the observational data of the neutrino annihilation energy $E_{\nu\bar{\nu}}$ for LGRBs. The LGRB samples come from the data in \citet{Nemmen2012}. Some GRBs with their durations being less than 2 s or longer than 300 s are not considered \citep{Leng2014,Liu2015c}. Furthermore, GRB 020903 is also excluded in our sample owing to the uncertain opening angle. The different colored lines correspond to $\dot{m}=0.1,~0.2,~0.3,~0.5$, and $1$, and the different typed lines correspond to $a_i=0.5,~0.9$, and $0.99$ with given $m_i=3$ and the typical red shift $z=2$. All lines are truncated when $M_{\rm disk}=5 ~M_{\odot}$. The magenta filled circles represent the LGRB data. In this figure, more than half of the data are under our predicted lines, which means that the neutrino annihilation processes can power these LGRBs in the reasonable terms. Although adopting extreme initial conditions, some LGRB data are still above or exceed the range of our lines, which reveals the limitation of the mechanism. To interpret these observed data, the alternative magnetohydrodynamics processes may be required \citep[e.g.,][]{Liu2015b,Liu2015c}, such as the BZ jet from the accretion system, the millisecond magnetar, or other processes \citep[e.g.,][]{Liu2012a,Liu2014,Liu2015a,Yuan2012}.

Finally, as shown in Figure 6, we also study the distribution of the disk mass in LGRBs and in SGRBs. We can easily find that the disk mass of more than half of LGRBs is below $5 ~M_{\odot}$, and others are beyond the limits, especially in the case of the lower BH spin parameter. Even for the case of Figure~6(b), there still exist some GRBs whose disk mass is larger than $11~M_{\odot}$. We can definitely eliminate the neutrino annihilation process in the center of those GRBs, which also means that the magnetohydrodynamics processes of the disk or magnetars should be considered.

\section{Conclusions and discussion}

We present the time-dependent mean neutrino annihilation luminosity and show its evolution with the different mean accretion rate and initial BH parameters. The possibilities of whether this model can account for the prompt emission and afterglow phases of GRBs are visited by comparing with the observations. The results show that our model can interpret most of data except for those of some high-energy LGRBs, which may require alternative magnetohydrodynamics processes in these GRBs. We further give the distribution of the disk mass by combining our model, fireball model,  and GRB samples. The results show that even for the extreme BH initial spins, there are still some GRBs requiring the massive disks, which approach or exceed the limits in simulations.

Gamma-ray prompt emission is followed by the early X-ray afterglow lasting from minutes to hours. X-ray flares are revealed by \emph{Swift} \citep{Gehrels2004} as a common feature of GRB afterglows \citep{Burrows2005}. They are frequently observed in LGRBs ($\sim30\%$; \citealt{Chincarini2007}) and, to a less extent, in SGRBs \citep{Barthelmy2005}. The most popular interpretation of X-ray flares is related to the restart of a GRB central engine \citep[e.g.,][]{Burrows2005,Zhang2006,Liu2008,Luo2013,Hou2014}. \citet{Perna2006} suggested that the flaring activities are observed in both LGRBs and SGRBs, and their observational properties are qualitatively consistent with the viscous disk evolution, so the origin of flares is considered as a hyperaccretion disk around a stellar-mass BH, as well as prompt emissions. They suggested that the gravitational instability occurring in the outer region of the disk may be the candidate for this variability. But \citet{Luo2013} argued that even considering the effects of the magnetic coupling and massive remnants of the disk, neutrino annihilation processes encounter difficulty in interpreting the X-ray flares lasting more than 100 s. This means that the BZ mechanism in the disk model or magnetar model is the better candidate for GRBs accompanied by the long-lasting X-ray flares. In consideration of the drastic X-ray flares observed in LGRBs, including GRB 050724 \citep[e.g.,][]{Dai2006,Falcone2007,Margutti2011}, the more extreme conditions are required in our model, and the range of application of neutrino annihilation processes may be reduced especially for LGRBs.

Furthermore, the afterglows of GRBs exhibit the abundant behaviors in the X-ray bands by analyzing the \emph{Swift} XRT data \citep[e.g.,][]{Zhang2006}. The unusually long `plateau' phases (shallow decay phases) lasting more than thousands of seconds are observed up to several hundred seconds after the burst trigger, and their plausible explanation is considered as the energy injection from a millisecond magnetar \citep[e.g.,][]{Dai1998,Fan2006b,Zhang2006}, and it is difficult to achieve by the neutrino annihilation processes. Actually, the plateau phases were observed in some GRBs in our sample, for example, GRB 070125, GRB 050904, GRB 090313, and GRB 090323 \citep[e.g.,][]{Kann2010}. So taking the plateau phase energy into account, our results would be more restricted for LGRBs.

It is worth noting that some physical conditions or mechanisms enhancing the neutrino annihilation luminosity could exist in this system. One of them is to increase the disk mass, the massive disks existing in binaries or collapsars is possible, especially for the binaries composed by $\sim 2~M_\odot$ NSs \citep[e.g.,][]{Demorest2010,Antoniadis2013,Strader2015}. However, the outflow from the NDAF can cause a mass loss and influence the total annihilation luminosity \citep[e.g.,][]{Liu2012b,Janiuk2013}. Fortunately, the outflow may appear only for the very low accretion rates when the advective cooling cannot balance the viscous heating \citep[e.g.,][]{Liu2008,Gu2015}. On the other hand, \citet{Liu2015a} showed the effects of the vertical convection on the structure and luminosity of the NDAF around a stellar-mass BH in spherical coordinates. They found that the neutrino luminosity and annihilation luminosity have a certain increase by considering the vertical convection to suppress the advection in the disk.

In recent years, the NDAF model as the central engine of GRBs was investigated by a lot of full hydrodynamic and magnetohydrodynamic (MHD) simulations \citep[e.g.,][]{Janka1999,Aloy2005,Burrows2007,Shibata2007,Harikae2010,Sekiguchi2011,Janiuk2013,Foucart2014,Just2015,Richers2015}. In the context of general relativity, \citet{Shibata2007} focused on the neutrino cooling of the accreting torus around the BH and the capture of the neutrino-trapping effect in a qualitative method. \citet{Janiuk2013} calculated the structure and neutrino emission of a two-dimensional, relativistic models and compared them with the results of one-dimensional simulations \citep{Janiuk2007}. They concluded that the neutrino luminosity may exceed the BZ jet luminosity and the subsequent neutrino annihilation will provide an additional source of power to the GRB emission. A BH torus evolution simulation was presented by \citet{Just2015}, containing the analysis of the neutrino and viscously driven outflows. The neutrino luminosity drops by orders of magnitude within tenths of a second in their model. The discrepancy may be due to variant parameters and methods of different models. The more reasonable and self-consistent numerical solutions of the NDAF model should be developed in the future to carefully compare with the GRB observations.

\begin{acknowledgements}
We thank the anonymous referee for very useful suggestions and comments. This work was supported by the National Basic Research Program of China (973 Program) under grant 2014CB845800, the National Natural Science Foundation of China under grants 11222328, 11233006, 11274200, 11333004, 11373002, 11473022, 11503011, and U1331101, and the Key Scientific Research Project in Universities of Henan Province under grant 15A160001.
\end{acknowledgements}

\clearpage

\begin{figure*}
\centering
\includegraphics[angle=0,scale=0.7]{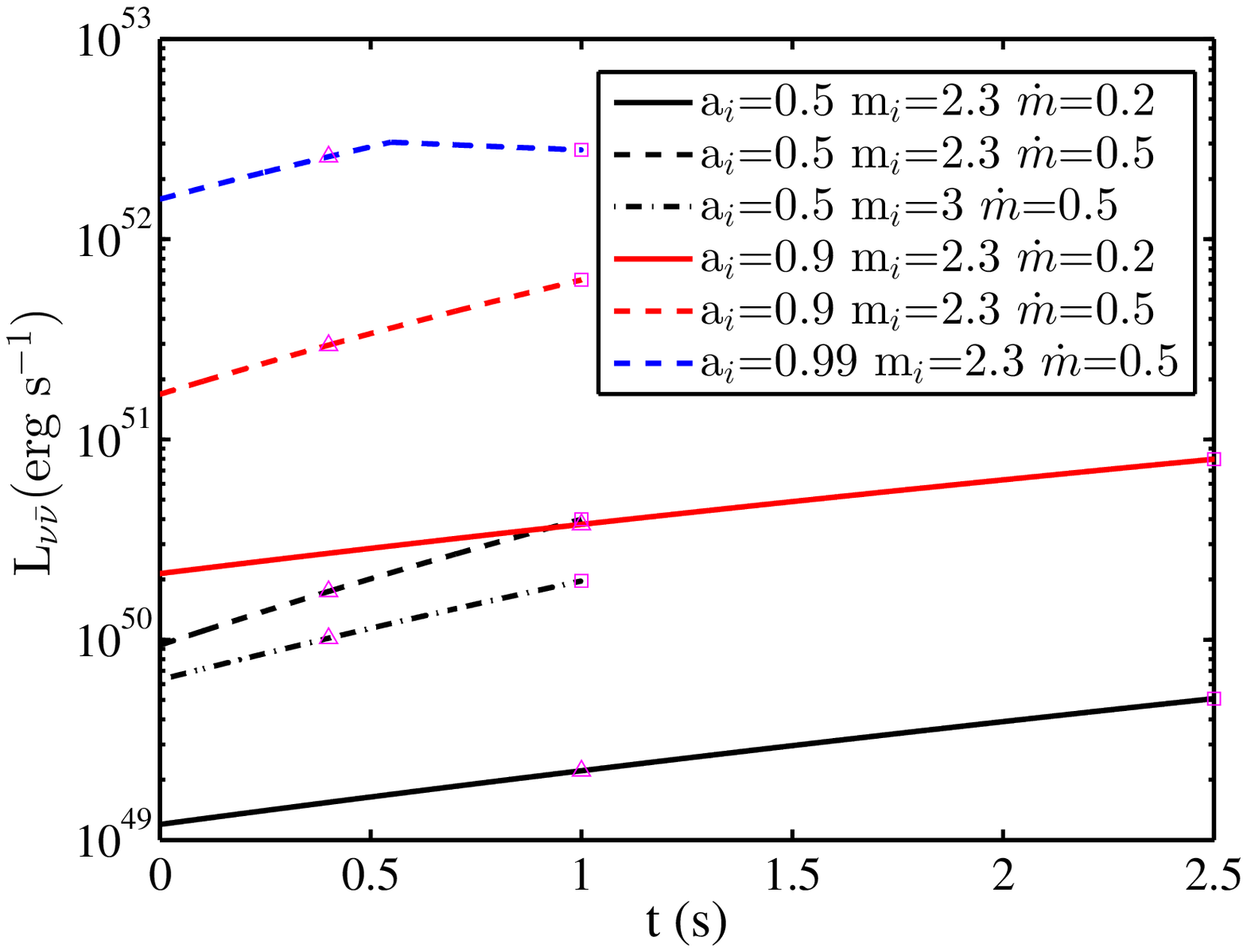}
\caption{Evolution of SGRBs' mean neutrino annihilation luminosity $L_{\nu\bar{\nu}}$. The black, red, and blue lines show the profiles with an initial spin of BH $a_i=0.5, 0.9,$ and $0.99$, respectively. The solid and dashed lines correspond to the dimensionless mean accretion rate $\dot{m}=0.2$ and $0.5$ with the initial dimensionless BH mass $m_{i}=2.3$, apart from the dot-dashed line with $m_{i}=3$. The disk masses $m_{\rm disk}$ are equal to $0.2,$ and $0.5~M_{\odot}$, which are marked with triangles and squares, respectively.}
\label{fig1}
\end{figure*}

\clearpage

\begin{figure*}
\centering
\includegraphics[angle=0,scale=0.7]{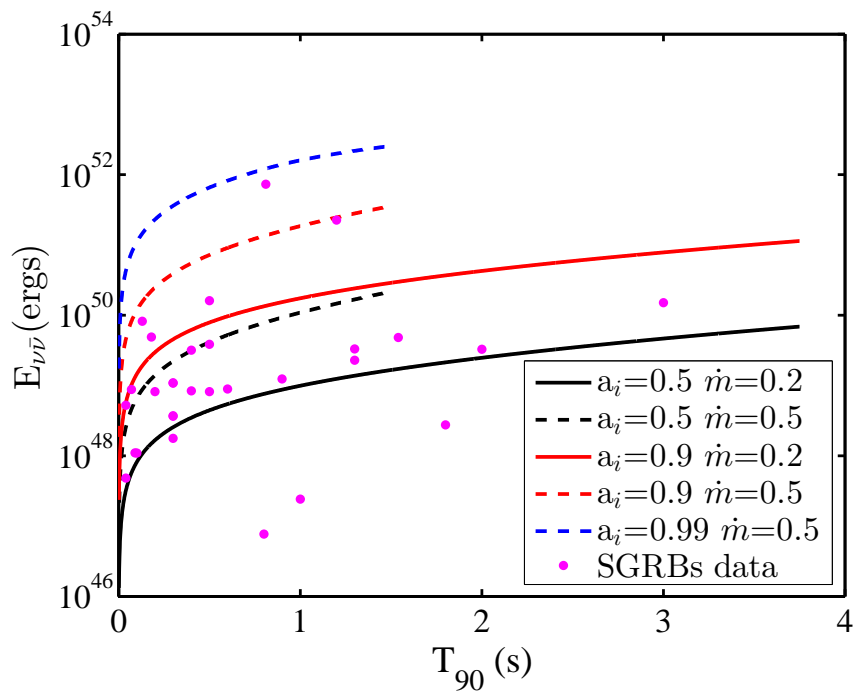}
\caption{Predictions of our model compared with SGRB observational data. The colors and styles of lines are the same as in Figure 1. All curves correspond to the initial dimensionless value of BH mass $m_i=2.3$ and typical red shift $z=0.5$. While disk mass $M_{\rm disk}=0.5~ M_{\odot}$, all lines are truncated. The magenta points show the neutrino annihilation energy calculated by observational data.}
\label{fig2}
\end{figure*}

\clearpage

\begin{figure*}
\centering
\includegraphics[angle=0,scale=0.5]{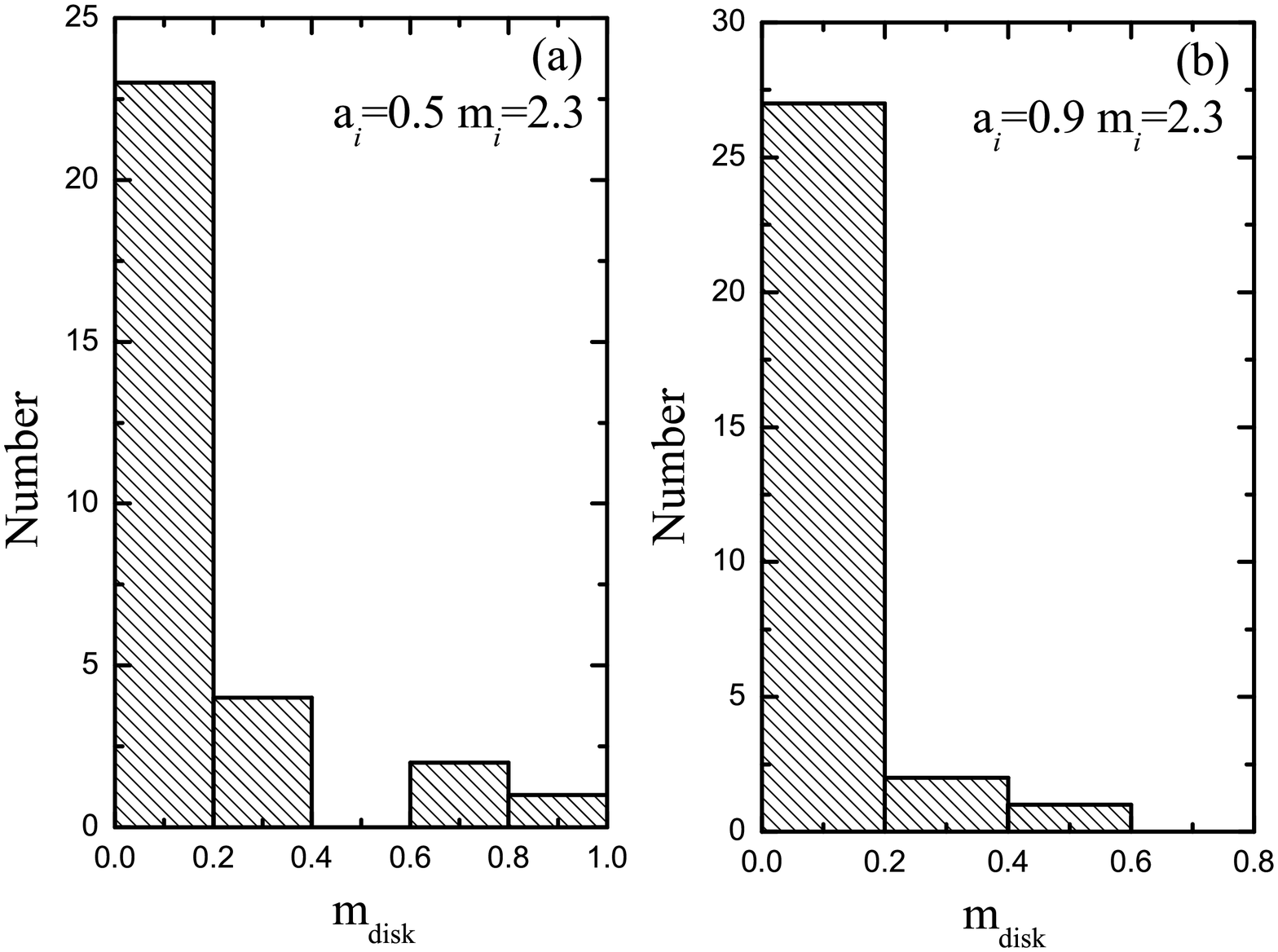}
\caption{Distribution of SGRB disk masses for different BH initial spins and the same initial BH mass $m_i=2.3$.}
\label{fig3}
\end{figure*}

\clearpage

\begin{figure*}
\centering
\includegraphics[angle=0,scale=0.7]{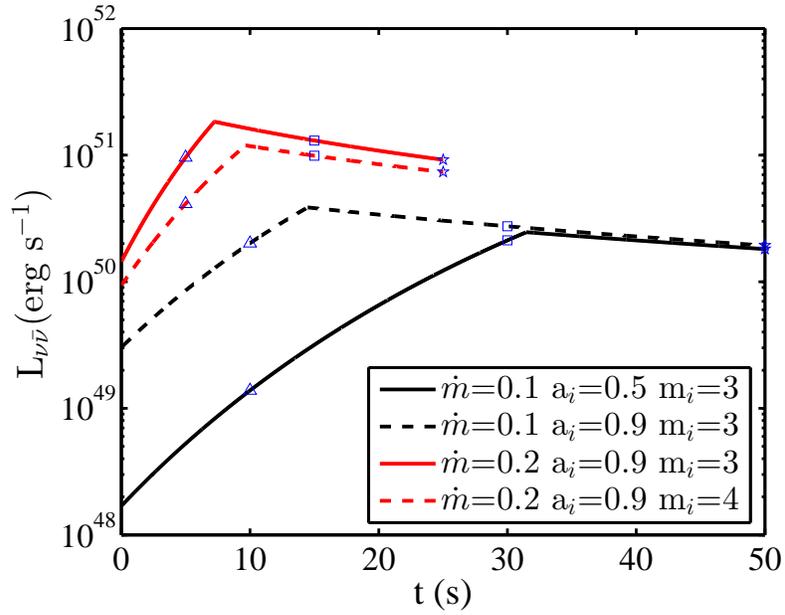}
\caption{Evolution of mean neutrino annihilation luminosity $L_{\nu\bar{\nu}}$ of LGRBs. The black solid and dashed lines correspond to the BH spin parameter $a_i=0.5$ and $0.9$ with $\dot{m}=0.1$ and $m_{i}=3$. The red solid and dashed lines correspond to $m_{i}=3$ and $4$ with $a=0.9$ and $\dot{m}=0.2$. $m_{\rm disk}$=1, 3, and 5 are marked with the triangles, squares, and five-pointed stars, respectively.}
\label{fig4}
\end{figure*}

\clearpage

\begin{figure*}
\centering
\includegraphics[angle=0,scale=0.7]{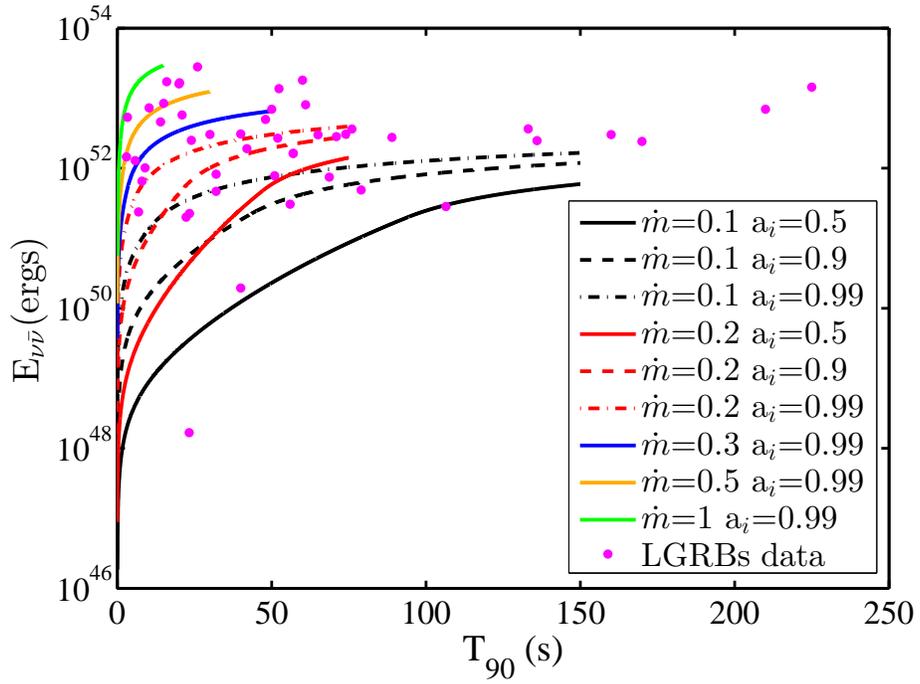}
\caption{Predictions of our model compared with LGRB observational data. The different colored lines corresponds to $\dot{m}$=0.1,~0.2,~0.3,~0.5,  and $1$; the different typed lines correspond to $a_i=0.5,~0.9$ and $0.99$ with given $m_i=3$ and the typical red shift $z=2$. All lines are truncated when $M_{\rm disk}=5 M_{\odot}$. The magenta filled circles represent the LGRB data.}
\label{fig5}
\end{figure*}

\clearpage

\begin{figure}
\centering
\includegraphics[angle=0,scale=0.5]{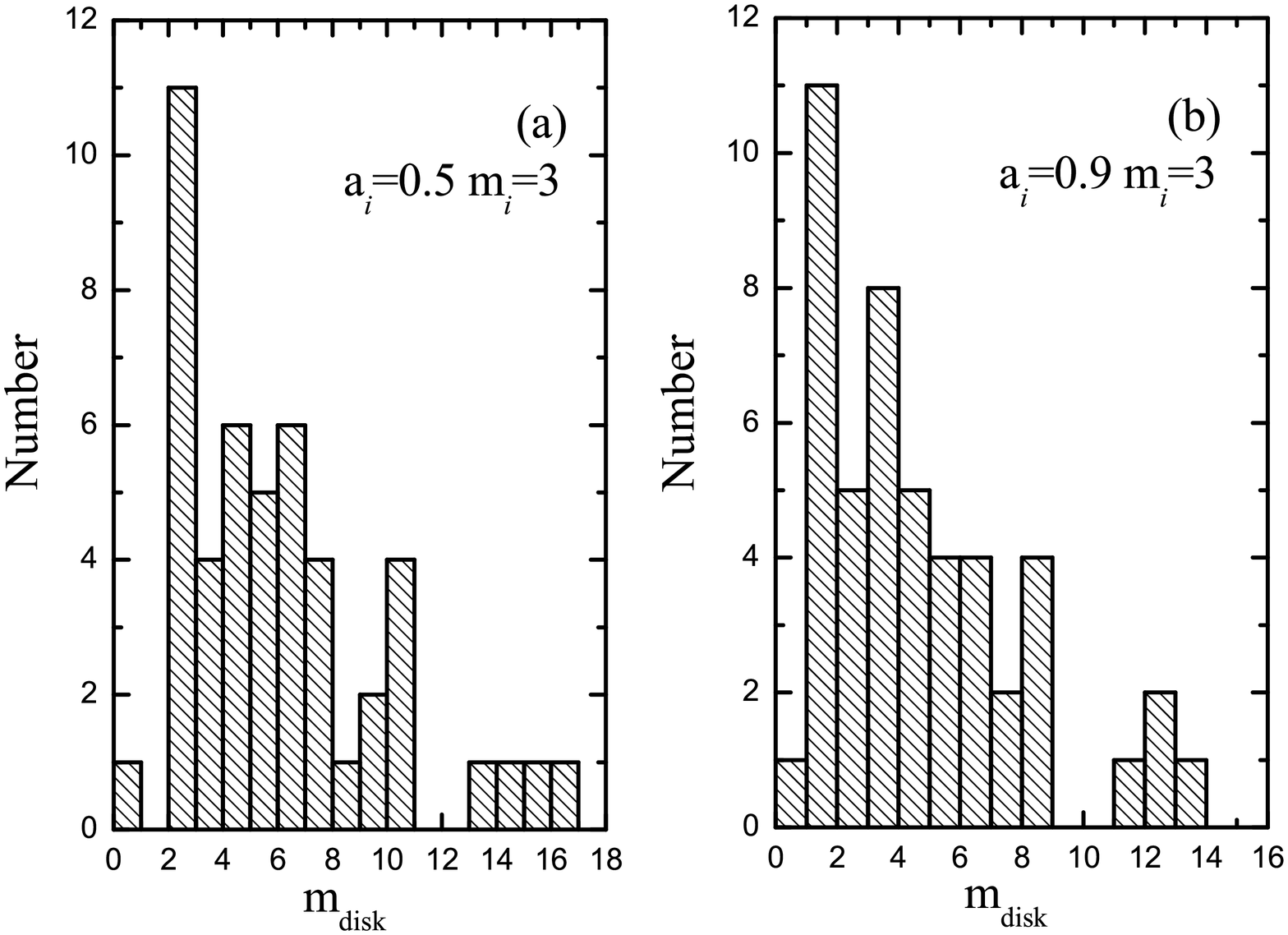}
\caption{Distribution of LGRBs disk masses for different BH initial spins and the same initial BH mass $m_{i}=3$.}
\label{fig6}
\end{figure}

\clearpage

\end{document}